\documentclass[11pt]{article}
\usepackage{graphicx}
\usepackage{epsfig}
\usepackage[figuresright]{rotating}
\usepackage{amsmath}
\usepackage{amssymb}
\usepackage{amsfonts}

\newcommand{\pibf}{\mbox{\boldmath $\pi$}}

\newcommand{\vectau}{\mbox{\boldmath$\tau$}}

\newcommand{\fdagger}{\mbox{$/\!\!\!\partial$}}
\usepackage{cite}
\begin{document}
\begin{center}
{\LARGE{\bf Electromagnetic polarizabilities of the nucleon and properties
of the $\sigma$-meson pole contribution}}\\[1ex]
Martin Schumacher\\mschuma3@gwdg.de\\
\renewcommand{\thefootnote}{\arabic{\footnote}}
Zweites Physikalisches Institut der Universit\"at G\"ottingen,
Friedrich-Hund-Platz 1\\ D-37077 G\"ottingen, Germany
\end{center}
\begin{abstract}
The $t$-channel contribution to the difference of 
electromagnetic polarizabilities of the nucleon, $(\alpha-\beta)^t$, 
 can be quantitatively understood in terms of a $\sigma$-meson pole in the 
complex $t$-plane of the invariant scattering amplitude $A_1(s,t)$
with properties of the  $\sigma$ meson as given by  the quark-level
Nambu--Jona-Lasinio model (NJL). Equivalently, this quantity
may be understood  in terms of a cut
in the complex $t$-plane where the properties of the $\sigma$ meson
are taken from the $\pi\pi\to\sigma\to\pi\pi$, $\gamma\gamma\to\sigma\to\pi\pi$
and $N\bar{N}\to \sigma \to \pi\pi$ reactions. 
This equivalence  may be understood as a sum rule
where the properties of the  $\sigma$ meson as predicted by the
NJL model are related 
to the $f_0(600)$ particle observed in the three   reactions. 
In the following we describe details of the derivation of
$(\alpha-\beta)^t$ making use of predictions of the quark-level NJL model
for the $\sigma$-meson mass.
\end{abstract}


\section{Introduction}

The $\sigma$ meson is an indispensable supplement of the pion 
\cite{schwinger57}. In terms of non-strange  quarks, the argument
 is that  four $q\bar{q}$
states should correspond to four mesons where the neutral members
of the $(\pibf,\sigma)$ isospin quartet have the flavor structures
$|\pi^0\rangle=(u\bar{u}-d\bar{d})/{\sqrt{2}}$ and 
$|\sigma\rangle=(u\bar{u}+d\bar{d})/{\sqrt{2}}$.
According to its flavor structure the $\sigma$ meson is a scalar-isoscalar
particle with relative angular momentum $L=1$ and spin $S=1$ of the two
quarks coupled to $J=0$. Therefore, it has the quantum numbers of
the vacuum  and, correspondingly, the $\sigma$-field entering into the linear
$\sigma$ model (L$\sigma$M) has 
a nonzero vacuum expectation value 
$\langle 0|\sigma|0\rangle \neq 0$. This leads to a 
mass $m_\sigma$ of the $\sigma$ meson which is not quantitatively predicted
by the L$\sigma$M, but quite naturally follows from 
the quark-level Nambu--Jona-Lasinio
model (NJL), by adjusting the predictions of this model to the pion decay
constant $f_\pi$, the average current-quark mass $m_0=\frac12 (m_u+m_d)$ 
and to the 
pion mass $m_\pi$. In this way parameters can be  avoided  
which otherwise would not be precisely determined. 
On the other hand, the $\sigma$ meson showing up as a broad
resonant intermediate state in reactions in which two pions 
are involved \cite{eidelman04} 
may be understood as a $(u\bar{u}+d\bar{d})/{\sqrt{2}}$,
$1^3P_0$ core state in a confining potential,  coupled to a 
$(\pi^+\pi^--\pi^0\pi^0+\pi^-\pi^+)/{\sqrt{3}}$ di-pion 
state in the continuum, where the two pions are in a relative $S$-state
with isospin $I=0$. 
This coupling lowers the average mass as compared to the confined core state 
and leads to a broad mass distribution.
This dual aspect of the $\sigma$ meson leads to two different predictions
for the $t$-channel contribution, $(\alpha-\beta)^t$, of the
difference of the electric and magnetic polarizability of the nucleon which
have been shown \cite{levchuk05} 
to agree with each other and with the experimental result. 
In the present work we give details of the
derivation of $(\alpha-\beta)^t$ making use of predictions of the  NJL model
for the $\sigma$-meson mass.

\section{The quantity $(\alpha-\beta)^t$ predicted from a   $\sigma$-meson
  pole }

Effective field theories are an excellent tool to adapt properties of QCD
to the low-energy regime. In applying these effective field theories
to phenomena like the polarizability of the nucleon, care has to be taken 
to find out what aspects of the phenomenon under consideration can be 
reproduced and where other theoretical tools are more appropriate.

\subsection{Outline of the problem and arguments in favor of the NJL
  model} 
\begin{figure}[ht]
\begin{center}
\includegraphics[width=0.5\linewidth]{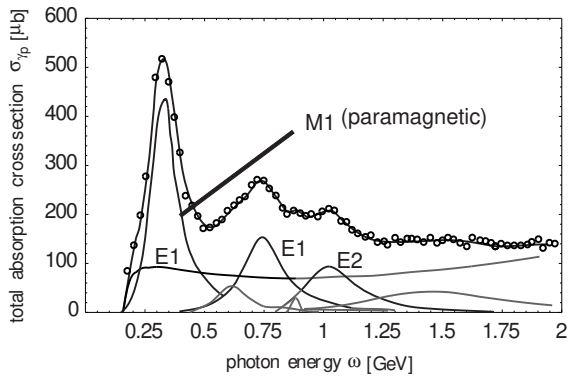}
\end{center}
\caption{ $s$-channel d.o.f.: Photoabsorption cross section separated
into multipoles} 
\label{s-channel}
\end{figure}

In case of Compton scattering 
\cite{lvov93a,lvov97,babusci98,drechsel03,wissmann04,schumacher05a,schumacher05b} two
types of degrees of
freedom (d.o.f.) of the nucleon have to be taken into account  which 
may  be termed $s$-channel d.o.f.  and $t$-channel d.o.f. 
The $s$-channel d.o.f. are those degrees of freedom of the nucleon which
also show up in photoabsorption experiments on the nucleon. The main
examples are the nonresonant $E_{0^+}$ channel of pion photoproduction
 corresponding to the 
``pion cloud'', and the $\Delta$ resonance. The  nonresonant $E_{0^+}$ channel
is of $E1$ multipolarity and contributes only about $40\%$ of the electric
polarizability $\alpha$, in partial contradiction to the frequently stated
belief that the ``pion cloud'' dominates the electric polarizability.
The results obtained \cite{levchuk05} for the
$s$-channel contributions are
$\alpha^s_p=4.5\pm 0.5$ and $\alpha^s_n=5.1\pm 0.6$ (in units of  $10^{-4}{\rm
  fm}^3$) for the proton and neutron, respectively, to be compared with the
corresponding experimental values 
$\alpha^{\rm exp}_p=12.0\pm 0.6$ and 
$\alpha^{\rm exp}_n=12.5\pm 1.7.$
The $\Delta$ resonance is the origin of  the  by far dominating part
of  the
electromagnetic polarizabilities. It is of $M1$ multipolarity and, therefore,
a strong source of paramagnetic polarizability. Here the numbers are
$\beta^s_p=9.5\pm 0.5$ and 
$\beta^s_n=10.1\pm 0.6$ to be compared with 
$\beta^{\rm exp}_p=1.9\pm 0.6$ and $\beta^{\rm exp}_n=2.7 \pm
1.8$. Apparently, there exists a strong diamagnetic polarizability which
cannot have its origin from  the $s$-channel d.o.f.
For illustration, the
$s$-channel d.o.f. are shown in Fig. \ref{s-channel}.
We see the strong $P_{33}(1232)$ ($\Delta$) resonance line and two other 
prominent lines corresponding to the $D_{13}(1520)$ and the $F_{15}(1680)$
resonances which are sources of $E1$ and $E2$ strength, respectively.
The main source of $E1$ strength is due to  the nonresonant part of the
cross section, which at the lower energies  is due to $1\pi$ photoproduction,
and due to $2\pi$ $\cdots$ photoproduction at higher energies.

The additional  $t$-channel
d.o.f. are required by Mandelstam analyticity \cite{hearn62}.   The invariant
amplitudes $A_i(s,t)$ are analytical functions of the two variable $s$
and $t$ and, therefore, the singularities of  the $t$-channel as well as 
those of the $s$-channel have to be taken into account.
The $t$-channel d.o.f.
may be identified with a  $(\pi^0,\sigma)$ doublet in the intermediate state
where the two mesons are coupled to two photons on the one side and to
constituent quarks on the other. In terms of linear polarization the two
cases differ from each other by the fact that for the $\pi^0$ meson the
directions of  linear polarization are perpendicular  whereas for 
the $\sigma$ meson they are parallel.
The $\pi^0$ meson corresponds to a pole in the $t$-plane which
contributes the dominant part, $\gamma^t_\pi$,
of  the backward 
spin polarizability $\gamma_\pi$.  The $\sigma$ meson contributes the dominant
part, $(\alpha-\beta)^t$, of the difference of electromagnetic
polarizabilities $(\alpha-\beta)$. As a broad mass distribution, the
$\sigma$ meson corresponds to a cut in the $t$-plane whereas as a particle
with a definite mass $m_\sigma$ it corresponds to a pole in analogy to the
$\pi^0$ meson case.

As far as effective field theories are concerned only those versions are of
interest which contain the $\sigma$ meson explicitly as a particle.
This means that the 
quark-level linear $\sigma$ model  (L$\sigma$M)
and the quark-level Nambu--Jona-Lasinio  model (NJL)
are candidates for representing the 
$t$-channel d.o.f. whereas the nonlinear $\sigma$ model and chiral
perturbation theory may be disregarded in connection with  the present purpose.
The Lagrangian of the L$\sigma$M consistently describes the mechanism
of spontaneous symmetry breaking but does not have the capability of
predicting the $\sigma$ meson mass $m_\sigma$ 
on an absolute scale \cite{gellmann60,dealfaro73}. 
This is different in the NJL model
\cite{nambu61,lurie64,eguchi76,vogl91,klevansky92,hatsuda94,bijnens96,thomas00}
which describes dynamical symmetry breaking and, thus, 
predicts a definite $\sigma$ meson mass, $m_\sigma$, 
instead of a broad mass distribution.
 This is the reason why we consider the  NJL model as the appropriate
candidate for our present study. However, the application of the NJL
should be restricted to predictions in connection with 
the $\sigma$-pole contribution. Applications to other aspects of the 
electromagnetic polarizabilities are not expected to lead to meaningful
results.

\subsection{The mass $m_\sigma$ of the $\sigma$ meson \label{subsection2-2}}

The Lagrangian of the NJL model 
has been formulated in two equivalent ways 
\cite{lurie64,eguchi76,vogl91,klevansky92}
\begin{eqnarray}
&&{\cal L}_{\rm NJL}=\bar{\psi}(i\fdagger-m_0) \psi
+ \frac{G}{2}[(\bar{\psi}\psi)^2+(\bar{\psi}i\gamma_5\vectau\psi)^2],
\label{NJL1}\\
{\rm and}\hspace{1cm} && \nonumber\\
&&{\cal L'}_ {\rm
  NJL}=\bar{\psi}i\fdagger\psi-g\bar{\psi}(\sigma+i\gamma_ 5
\vectau\cdot\pibf)\psi-\frac12\delta\mu^2(\sigma^2+\pibf^2)+\frac{gm_0}{G}
\sigma,
\label{NJL2}\\
{\rm where} \hspace{1cm}&&\nonumber\\
&&G=g^2/\delta\mu^2\quad \mbox{and}\quad \delta\mu^2=(m^{\rm cl}_\sigma)^2.
\label{grelations}
\end{eqnarray}
\begin{figure}[ht]
\begin{center}
\includegraphics[width=0.5\linewidth]{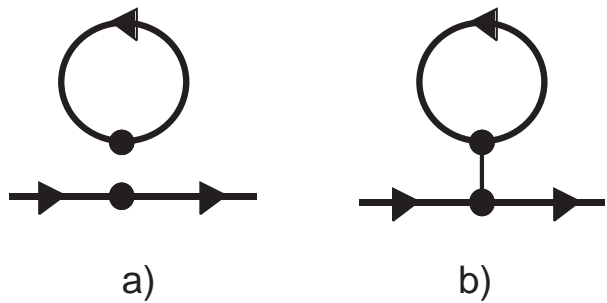}
\end{center}
\caption{a) Four-fermion theory tadpole diagram, b) bosonized tadpole
  diagram \cite{lurie64}.}
\label{tadpolegraphs}
\end{figure}
Eq. (\ref{NJL1}) describes the four-fermion version of the
NJL model and eq.  (\ref{NJL2}) the bosonized  version. 
The quantity $\psi$ denotes the spinor of constituent quarks with two
flavors. The quantity $G$ is the coupling constant of the four-fermion
version, $g$ the Yukawa coupling constant and $\delta\mu$ a mass parameter
entering into the mass counter-term of eq. (\ref{NJL2}). The coupling
constants $G$, $g$ and  the mass parameter $\delta\mu$ 
are related to each other 
and to the $\sigma$ meson mass in the chiral limit (cl),
$m^{\rm cl}_\sigma$, as
given in (\ref{grelations}).  The relation $\delta\mu^2=(m^{\rm
    cl}_\sigma)^2$ can easily be derived by applying spontaneous symmetry
  breaking to $\delta\mu$ in analogy 
to spontaneous symmetry breaking predicted by the
  L$\sigma$M for the mass parameter $\mu$ entering into this model
\cite{dealfaro73}.
In the chiral limit  ($m_0\to 0$) the version of Eq. (\ref{NJL2}) 
contains the spinor-dependent term
in the same way as the  L$\sigma$M, whereas  the major part of the  bosonic
$(\pibf,\sigma)$ dependent terms of the L$\sigma$M are absent. 
This
means that with respect to the spinor dependent terms the 
L$\sigma$M and the NJL model are equivalent whereas only a truncated
version of the bosonic part of the L$\sigma$M is produced. 
The terms in Eqs.  (\ref{NJL1}) and (\ref{NJL2}) containing the average 
current-quark mass $m_0$ describe explicit symmetry breaking and can be shown
to be equivalent.
The insight that the L$\sigma$M and the NJL model are basically
equivalent dates back to the 1960s and 1970s. Fig. \ref{tadpolegraphs} shows
the four-fermion theory tadpole diagram and the bosonized  tadpole diagram.
The underlying idea has been worked
out in detail by Eguchi \cite{eguchi76}, by Vogl and Weise \cite{vogl91}
and Delbourgo and Scadron \cite{delbourgo95}.
Both versions
can be exploited to make a prediction for $m_\sigma$ on an absolute scale.
The NJL model faces the problem that use is made of integrals in momentum
space which diverge in the infinite momentum limit. To overcome this
problem two different regularization schemes have been developed.  
The regularization through a cut-off momentum $\Lambda$ restricts the
evaluation of the integrals to the low-momentum region. 
Dimensional regularization treats the integrals without a cut off
but makes use of the fact that the difference between two diverging integrals
is finite.

\subsubsection{Four-fermion version of the NJL model with 
regularization through a
  cut-off parameter $\Lambda$}
Using diagrammatic techniques the following  equation may be found 
\cite{klevansky92,hatsuda94}
\begin{eqnarray}
&&M^*=m_0+ 8\, i\, G N_c \int^{\Lambda}\frac{d^4 p}{(2\pi)^4}
\frac{M^*}{p^2-M^{*2}},
\label{gapdiagram}\\
&&f^2_\pi = -4\,i\,N_cM^{*2} \int^{\Lambda}\frac{d^4p}{
(2\pi)^4}\frac{1}{(p^2-M^{*2})^2},
\label{fpiexpress}\\
&&m^2_\pi=-\frac{m_0}{M^*}\frac{1}{4\,i\,G N_c I(m^2_\pi)},\quad
I(k^2)=\int^{\Lambda}\frac{d^4p}{(2\pi)^4}\frac{1}{[(p+\frac12 k)^2-M^{*2}][
(p-\frac12 k)^2-M^{*2}]}.
\label{pionmass-2}
\end{eqnarray}
The expression given in (\ref{gapdiagram}) is the gap equation with $M^*$
being the mass of the constituent quark with the contribution 
$m_0=(m _u+m_d)/2$ of the
current
quarks included and $N_c=3$ being the number of colors. Eq.
(\ref{fpiexpress}) represents  the pion decay constant having 
the experimental value $f_\pi=(92.42\pm 0.26)$  MeV \cite{eidelman04}.
The expression given in (\ref{pionmass-2}) is equivalent 
to  the Gell-Mann--Oakes--Renner relation \cite{gellmann68}
when formulating (\ref{pionmass-2}) in the chiral limit.
Using  (\ref{gapdiagram})  -- (\ref{pionmass-2}) it
is possible to calculate the quantities $M^*$, $f_\pi$ and
$m_\pi$ simultaneously and to adjust the parameters $G$ and $\Lambda$
in such a way that the experimental values for $f_\pi$, $m_\pi$ and $m_0$ are 
reproduced. It is apparent that this procedure leaves no room for an unknown
parameter so that the predicted value for $m_\sigma$ is model independent
except, of course, for the general use of the theoretical frame
provided by the  NJL model. Numerical calculations of the type outlined 
above have been carried out by Hatsuda and Kunihiro \cite{hatsuda94}
applying RPA techniques.
The result obtained is
\begin{equation}
m_\sigma \simeq 668 \,\,\,{\rm MeV}
\label{numericalresult-2}
\end{equation} 
where $m_\sigma\simeq 2M^*$ has been  applied.

\subsubsection{Bosonized NJL model or dynamical L$\sigma$M with dimensional
regularization}

A second way to predict $m_\sigma$ on an absolute scale introduced by
Delbourgo and Scadron  \cite{delbourgo95}          is obtained by
exploiting Eqs. (\ref{NJL2}) and (\ref{grelations}).
The starting point \cite{delbourgo95}
are  representations of the pion decay constant
and the gap equation for the constituent-quark mass, M, in the chiral limit 
\begin{eqnarray}
&&f^{\rm cl}_\pi=-4iN_cgM\int \frac{d^4p}{(2\pi)^4}\frac{1}{(p^2-M^2)^2},
\label{fpicl}\\
&&M=-\frac{8 i N_c g^2}{(m^{\rm cl}_\sigma)^2}
\int\frac{d^4 p }{(2\pi)^4}\frac{M}{p^2-M^2}. 
\label{LSMandNJL-2}
\end{eqnarray}
The third equation, (\ref{pionmass-2}),  
for the pion mass $m_\pi$ is not needed because this
quantity is equal to zero in the chiral limit, $m^{\rm cl}_\pi=0$.
Except for a strict application of the chiral limit, these 
equations differ 
from those in (\ref{gapdiagram}) and (\ref{fpiexpress}) by the fact
that use has been made
of the Goldberger-Treiman relation for the chiral limit
\begin{equation}
gf^{\rm cl}_\pi=M
\label{goldtrei}
\end{equation}
in (\ref{fpicl}) and by replacing $G$ by $-g^2/(m^{\rm cl}_\sigma)^2$  in
(\ref{LSMandNJL-2}).
The different signs in front of the integrals in 
(\ref{gapdiagram}) and (\ref{LSMandNJL-2}) follow from the fact that different
regularization schemes are applied in the two cases \cite{delbourgo95}.

Then, making use of the identity (dimensional regularization 
\cite{thomas00,delbourgo95})
\begin{equation}
-i\frac{(m^{\rm cl}_\sigma)^2}{16N_cg^2}=
\int\frac{d^4 p}{(2\pi)^4}\left[\frac{M^2}{(p^2-M^2)^2}-
\frac{1}{p^2-M^2}\right]
=-\frac{iM^2}{(4\pi)^2}
\label{identity}
\end{equation}
we arrive at
\begin{equation}
(m^{\rm cl}_\sigma)^2=\frac{N_c g^2 M^2}{\pi^2},
\label{sigmamasssquared}
\end{equation}
and with $m^{\rm cl}_\sigma=2M$ at
\begin{equation}
g=g_{\pi qq}=g_{\sigma qq}= 2\pi/\sqrt{N_c}=3.63.
\label{coupling}
\end{equation}
The $\sigma$-meson mass corresponding to this coupling constant is
\begin{equation}
m_\sigma= 666.0\,\,\,{\rm MeV},
\label{sigmamassfinal-2}
\end{equation}
where use has been made of
$m^2_\sigma= (m^{\rm cl}_\sigma)^2 + m^2_\pi$,
$f^{\rm cl}_\pi=89.8$ MeV \cite{nagy04}, $M=325.8$ MeV  and $m_\pi= 138.0$ MeV.
It is satisfying  to note that the results given in 
(\ref{numericalresult-2}) and (\ref{sigmamassfinal-2}) are in an excellent
agreement with each other. Since the procedures to arrive at $m_\sigma$
are quite different in the two cases the good agreement of the two results
gives us confidence that value obtained for $m_\sigma$ is on a stable basis.

\subsection{The two-photon decays of the 
$f_0(980)$, the $\pi^0$ and the $\sigma$
meson \label{Thetwo-photon}}

For mesons $P$ having the constituent-quark structure
\begin{equation}
|q\bar{q}\rangle=\frac{a|u\bar{u}\rangle + b  |d\bar{d}\rangle
+ c  |s\bar{s}\rangle}{\sqrt{a^2+b^2+c^2}}
\label{quarkstructure}
\end{equation}
the two-photon amplitude is given in  the generic form 
\begin{equation}
|M(P\to \gamma\gamma)|= \frac{\alpha_e}{\pi f_P} 
N_c\,\sqrt{2}\,\frac{a\, e^2_u+ b\, e^2_d 
+ c\,(\hat{m}/m_s)\,e^2_s}{\sqrt{a^2+b^2+c^2}},
\label{twophotonamplitude}
\end{equation}
where $\alpha_e=1/137.04$, $f_P$ the decay constant of the meson $P$
(see e.g. \cite{donoghue85,cooper88}) and $\hat{m}/m_s\approx 1/\sqrt{2}$ the
ratio of light and strange constituent quark masses. However, we can use 
\begin{equation}
f_P \simeq f_\pi
\label{pidecayconst}
\end{equation}
without a major loss of precision. The reason is that $f_P$ does 
not depend on the
flavor wave function of the meson as can be seen in 
(\ref{gapdiagram}). Small deviations from 
(\ref{pidecayconst}) only occur when there is a strange-quark content in the
meson as in case of $\eta$, $\eta'$ and $f_0(980)$, 
because of the larger current-quark mass of the strange quark.
This leads us to the following
relations
\begin{eqnarray}
&|M(f_0(980)\to\gamma\gamma)|&=\frac{\alpha_e}{\pi f_\pi}N_c \left[
  \left(-\frac13\right)^2  \right]=\frac13
\frac{\alpha_e}{\pi f_\pi},\label{width2}\\
&|M(\sigma\to\gamma\gamma)|&=\frac{\alpha_e}{\pi f_\pi}N_c \left[
\left(\frac23\right)^2 +  \left(-\frac13\right)^2  \right]=\frac53
\frac{\alpha_e}{\pi f_\pi},\label{width3}\\
&|M(\pi^0\to\gamma\gamma)|&=\frac{\alpha_e}{\pi f_\pi}N_c \left[
\left(\frac23\right)^2 -  \left(-\frac13\right)^2  \right]=
\frac{\alpha_e}{\pi f_\pi},\label{width4}\\
&\Gamma_{f_0(980)\to\gamma\gamma}&=\frac{m^3_{f_0(980)}}{64\pi}
|M(f_0(980)\to\gamma\gamma)|^2= 0.33\,\, {\rm keV},\label{width6}\\
&\Gamma_{\sigma\to\gamma\gamma}&=\frac{m^3_\sigma}{64\pi}
|M(\sigma\to\gamma\gamma)|^2=2.6\,\, {\rm keV},\label{width7}\\
&\Gamma_{\pi^0\to\gamma\gamma}&=\frac{m^3_{\pi^0}}{64\pi}
|M(\pi^0\to\gamma\gamma)|^2=7.73\times10^{-3}\,\, {\rm keV},\label{width8}
\end{eqnarray}
where $f_P=f_\pi$ is used in all three cases. 
For the $\sigma$ meson we use the
``model independent'' $\sigma$ mass, $m_\sigma=666$ MeV,  determined
in the previous subsection using the method of
\cite{delbourgo95}.
The pion decay constant is $f_\pi= (92.43\pm 0.26)$ MeV.
For the $\sigma$ meson we then obtain
$\Gamma_{\sigma\to\gamma\gamma}=2.6 \,\,\,\mbox{keV}$
which agrees with the experimental result of Boglione and Pennington 
\cite{boglione99} $\Gamma_ {\sigma\to\gamma\gamma}=(3.8\pm 1.5)$ keV
within the errors. 
In (\ref{width3}) it has been assumed
that the coupling of the two photons to the $\sigma$ meson 
proceeds only through its  non-strange quark content and that there are
no additional contributions due to meson loops as obtained in the frame
of the L$\sigma$M  \cite{beveren02,scadron04}. From a theoretical point of
view this neglect of meson-loop contributions is justified through 
the use of the NJL model where such additional contributions are absent.
For sake of completeness we note that for 
the $\pi^0$ meson the prediction is 
$\Gamma_{\pi^0\to\gamma\gamma}=7.73\times 10^{-3}$ keV to be compared with the
experimental value 
$\Gamma_{\pi^0\to\gamma\gamma}=(7.74\pm 0.55)\times 10^{-3}$ keV
obtained from the mean lifetime $\tau_{\pi^0}=(8.4\pm 0.6)\times 10^{-17}$ s
and the branching ratio $\pi^0\to\gamma\gamma$ of $(98.798\pm 0.032)\%$
\cite{eidelman04}.
We see that for the $\pi^0$ meson as well as for its chiral partner, the
$\sigma$ meson, the coupling to two photons may be understood as proceeding
through their $q\bar{q}$ internal  structures. 
It is remarkable to note that the prediction for the two-photon width of
$f_0(980)$ based on a $|\bar{s}s\rangle$ quark structure is in agreement 
with the experimental value
$\Gamma_{f_0(980)\to\gamma\gamma}=(0.39^{+0.10}_{-0.13} )$ keV given by
the particle data group PDG \cite{eidelman04} and also with 
$\Gamma_{f_0(980)\to\gamma\gamma}=(0.28^{+0.09}_{-0.13} )$ keV
given by Boglione and Pennington \cite{boglione99}. 

There was a
longstanding discussion whether or not the $q\bar{q}$ structure of scalar
mesons with masses below 1 GeV
should be replaced by a $(qq)(\bar{q}\bar{q})$ structure
\cite{jaffe76}. This alternative
appears to be strongly disfavored by the insight that scalar mesons 
neither correspond to  a $q\bar{q}$ structure nor to a 
$(qq)(\bar{q}\bar{q})$ structure, but to a  $q\bar{q}$ structure component
coupled to di-meson states
\cite{beveren86}.
Also, the good agreement of the two-photon decay width of the
$f_0(980)$ meson predicted on the basis
of a $|s\bar{s}\rangle$ structure with the corresponding experimental values
may be considered as an argument in favor of this structure.

\subsection{Polarizabilities and invariant amplitudes}

For the discussion of the polarizabilities of the nucleon 
in terms of Compton scattering the forward direction
($\theta=0$) and the backward direction ($\theta=\pi$) are of special
interest. Denoting the spin-independent and spin-dependent amplitudes
for the forward and backward direction by $f_0$, $g_0$, $f_\pi$ and
$g_\pi$, respectively, we arrive at
\begin{eqnarray}
&&f_0(\omega)= -\frac{\omega^2}{2\pi}\left[A_3(\nu,t)+ A_6(\nu,t)
\right],\quad\quad\quad
g_0(\omega)=\frac{\omega^3}{2\pi {m_N}}A_4(\nu,t), \label{T3}\\
&&f_\pi(\omega)=-\frac{\omega\omega'}{2\pi}\left(1+\frac{\omega\omega'}
{{m_N}^2}\right)^{1/2}\left[ 
A_1(\nu,t) - \frac{t}{4 {m_N}^2}A_5(\nu,t)\right],\label{T4}\\
&&g_\pi(\omega)=-\frac{\omega\omega'}{2\pi {m_N}}\sqrt{\omega\omega'}
\left[
A_2(\nu,t)+ \left(1-\frac{t}{4 {m_N}^2}\right)A_5(\nu,t)\right]
,\label{T5}\\
&&\omega'(\theta=\pi)=\frac{\omega}{1+2\frac{\omega}{{m_N}}},\,\,
\nu=\frac12 (\omega+\omega'),\,\, t(\theta=0)=0,
\,\,t(\theta=\pi)=-4\omega\omega',
\label{T6}
\end{eqnarray}
where $A_i$  
are the invariant amplitudes in the standard definition 
\cite{lvov93a,lvov97,babusci98,drechsel03,wissmann04,schumacher05a,schumacher05b}
and $m_N$ the nucleon mass.
For the electric, $\alpha$, and magnetic, $\beta$,  polarizabilities 
and the spin polarizabilities $\gamma_0$ and $\gamma_\pi$ for the
forward and backward directions, respectively, 
we obtain the relations
\begin{eqnarray}
&&\alpha+\beta = -\frac{1}{2\pi}\left[A^{\rm nB}_3(0,0)+ 
A^{\rm nB}_6(0,0)\right], \quad 
\alpha-\beta = -\frac{1}{2\pi}
\left[A^{\rm nB}_1(0,0)\right], \nonumber\\ 
&&\gamma_0= \frac{1}{2\pi {m_N}}\left[A^{\rm nB}_4(0,0)
\right], \quad\quad\quad\quad \quad\quad\quad\,\,\,
\gamma_\pi = -\frac{1}{2\pi {m_N}}
\left[A^{\rm nB}_2(0,0)+A^{\rm nB}_5(0,0) \right],
\label{T7}
\end{eqnarray}
where $A_i^{\rm nB}$ are the non-Born parts of the invariant amplitudes.

According to Eqs. (\ref{T3}) to  (\ref{T7}) the following linear
combinations of invariant amplitudes are of special importance because 
they contain the physics of the four fundamental sum rules, { \it viz.}
the BEFT \cite{bernabeu74} sum rule for $(\alpha-\beta)$, the
LN \cite{lvov99} sum rule for $\gamma_\pi$, the BL \cite{baldin60} sum rule
for $(\alpha+\beta)$, 
and the GDH \cite{gerasimov66} sum rule for the square of the
anomalous magnetic moment $\kappa^2$, respectively:
\begin{eqnarray}
&& {\tilde A}_1(\nu,t)\equiv A_1(\nu,t)-\frac{t}{4{m_N}^2}A_5(\nu,t),\label{T8}\\
&& {\tilde A}_2(\nu,t)\equiv A_2(\nu,t)+\left(1-\frac{t}{4{m_N}^2}\right)
A_5(\nu,t),\label{T9}\\
&& {\tilde A}_3(\nu,t)  \equiv A_{3+6}(\nu,t)\equiv A_3(\nu,t)+ 
A_6(\nu,t), \label{T10}\\
&& {\tilde A}_4(\nu,t)   \equiv A_4(\nu,t).
\label{T11}
\end{eqnarray}
Therefore, these linear combinations of invariant amplitudes may be considered
as generalized polarizabilities containing the essential physics of
the polarizability of the nucleon.

In the following use is made of the relations concerning $\alpha+\beta$,
$\alpha-\beta$ and $\gamma_\pi$, whereas the relation concerning $\gamma_0$
is written down only for the sake of completeness.

\subsection{Fixed-$\theta$ dispersion relations at $\theta=\pi$}

Fixed-$\theta$ dispersion relations have the advantage that a clear-cut
separation of the $s$-channel and the $t$-channel is possible, i.e.
there definitely is no double counting of empirical input \cite{hearn62}. 
In the following
we are interested in the backward spin-polarizability $\gamma_\pi$ and the
difference of electromagnetic polarizabilities $\alpha-\beta$ which have
a firm relation to the invariant amplitudes at $\theta=\pi$ where one  
can write down \cite{drechsel03,hoehler83}
a dispersion integral as
\begin{eqnarray}
{\rm Re}A_i(s,t)&=& A^{\rm B}_i(s,t)+A^{t-{\rm
    pole}}_i(s,t)\nonumber\\
&+& \frac{1}{\pi}\int^\infty_{({m_N}+m_\pi)^2} ds'{\rm
    Im}_s A_i(s',\tilde{t})
\left[\frac{1}{s'-s}+\frac{1}{s'-u}-\frac{1}{s'}\right]\nonumber\\
&+&\frac{1}{\pi} \int^\infty_{4m^2_\pi} dt' \frac{{\rm Im}_t
 A_i(\tilde{s},t')}{t'-t},
\label{Drechsel(185)}
\end{eqnarray}
where  ${\rm Im}_s A(s',\tilde{t})$
is evaluated along the hyperbola given by
\begin{equation}
s'u'={m_N}^4, \quad s'+\tilde{t}+u'=2{m_N}^2,
\label{Drechsel(186)}
\end{equation}
and ${\rm Im}_t A_i(\tilde{s},t')$  runs along the path defined by
the hyperbola
\begin{equation}
\tilde{s}\tilde{u}={m_N}^4, \quad \tilde{s}+t'+\tilde{u}=2{m_N}^2.
\label{Drechsel(187)}
\end{equation}
The amplitude $A^{t-{\rm pole}}_i(s,t)$ entering into
(\ref{Drechsel(185)})
describes the contribution of $t$-channel poles to the scattering amplitudes 
which for pseudoscalar mesons may be written in the form
\begin{equation}
A^{{\pi^0}+\eta+\eta'}_2(t)=\frac{g_{{\pi} NN}F_{\pi^0\gamma\gamma}}
{t-m^2_{\pi^0}}\tau_3+\frac{g_{{\eta} NN}F_{\eta\gamma\gamma}}
{t-m^2_{\eta}}+\frac{g_{{\eta'} NN}F_{\eta'\gamma\gamma}}
{t-m^2_{\eta'}},
\label{pseudoscalarPole}
\end{equation}
where the quantities $g_{\pi NN}$, {\it etc.} are the meson-nucleon coupling
constants and the quantities $F_{\pi^0 \gamma\gamma}$, {\it etc.} the 
two-photon decay amplitudes.
The last term in (\ref{Drechsel(185)}) represents the contribution of 
$t$-channel cuts to the scattering amplitudes which later on will be 
discussed in connection with the scalar-isoscalar 
$t$-channel. From (\ref{pseudoscalarPole})
the following relation for the $t$-channel part of the backward
spin-polarizability may be obtained
\begin{equation}
\gamma^t_\pi= \frac{1}{2\pi {m_N}}\left[\frac{g_{{\pi} NN}
F_{\pi^0\gamma\gamma}}
{m^2_{\pi^0}}\tau_3+\frac{g_{{\eta} NN}F_{\eta\gamma\gamma}}
{m^2_{\eta}}+\frac{g_{{\eta'} NN}F_{\eta'\gamma\gamma}}
{m^2_{\eta'}}\right].
\label{tchannelgammapi}
\end{equation}
The pion-nucleon coupling constant is given by the experimental value
$g_{\pi NN}= 13.169\pm 0.057$ \cite{bugg04}. 
The corresponding quantities for the
$\eta$ and $\eta'$ meson cannot be much different. The reason is that on a
  quark level and in the chiral limit the coupling constant $g_{\pi qq}$
shows up as a universal proportionality constant between the quantities 
$f^{\rm cl}_\pi$ and $M$ given in (\ref{fpicl}) and (\ref{LSMandNJL-2}),
respectively.  A different problem is the sign of the 
two-photon decay amplitude  where we have
\begin{equation}
F_{\pi^0\gamma\gamma}=-|M(\pi^0\to\gamma\gamma)|,\quad 
F_{\eta\gamma\gamma}=\pm|M(\eta\to\gamma\gamma)|,\quad
F_{\eta'\gamma\gamma}=\pm|M(\eta'\to\gamma\gamma)|.
\label{signsofamplitudes}
\end{equation}
The minus sign in case of $F_{\pi^0\gamma\gamma}$ is well established 
\cite{terentev73,lvov97,lvov99}
whereas the other signs are less well known.

The expression (\ref{tchannelgammapi})
for the $t$-channel part of the backward spin-polarizability
has been tested and found valid. Details may be found in Table \ref{spinpo}.
\begin{table}[h]
\caption{Backward spin-polarizability for the proton and the neutron (units
$10^{-4} {\rm fm}^4$)}
\begin{center}
\begin{tabular}{lllll}
\hline
1&Spin pol.& proton&neutron&\\
\hline
2&$\gamma_\pi$& $-38.7\pm 1.8$& $+58.6\pm 4.0$&Experiment 
\cite{schumacher05b}\\
3&$\gamma^s_\pi$& $+7.1 \pm 1.8$& $+9.1\pm 1.8$ & Sum rule\cite{lvov99}\\
\hline
4&$\gamma^t_\pi$&$-45.8\pm 2.5$&   $+49.5\pm 4.4$& line 2 -- line 3\\
5&$\gamma^t_\pi$&$-46.7$&$+46.7$&$\pi^0$-pole only\\
6&$\gamma^t_\pi$& $-45.1$& $+48.3$&$\pi^0+\eta+\eta'$-poles\,\,\mbox{a)}\\
7&$\gamma^t_\pi$& $-48.3$& $+45.1$&$\pi^0+\eta+\eta'$-poles\,\,\mbox{b)} \\
\hline
\end{tabular}\\[1ex]
a) $\eta$ and $\eta'$ contributions assumed to be positive numbers (line 6),\\ 
b) $\eta$ and $\eta'$ contributions assumed to be negative numbers (line 7)
\end{center}
\label{spinpo}
\end{table}
In \cite{lvov99} it is proposed to adopt the minus sign for the $\eta$ and 
$\eta'$ contributions in Eq. (\ref{signsofamplitudes}). By comparing
line 4 with lines 6 and 7 in Table \ref{spinpo} 
we see that a better agreement
with the experimental values is obtained when the plus sign is used.

For $(\alpha-\beta)$ we have the choice to either use     the pole
representation as appropriate for the  $\sigma$ meson as entering into
the NJL model or the cut
representation as appropriate for the $\sigma$ meson as a broad resonant
intermediate state. 

\begin{figure}[ht]
\begin{center}
\includegraphics[width=0.5\linewidth]{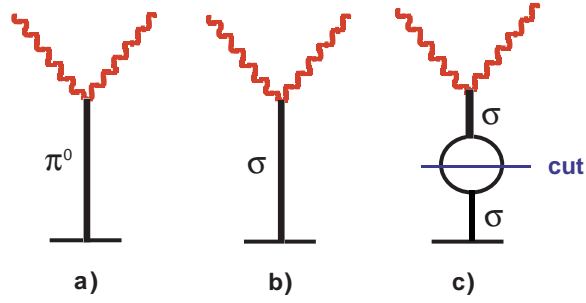}
\end{center}
\caption{a) $\pi^0$ pole diagram. b) $\sigma$ pole
  diagram, c) scalar-isoscalar $t$-channel as entering into the BEFT 
sum rule}
\label{poleandcutgraphs}
\end{figure}

Here we first
discuss the pole representation (see Fig. 3 b)).  
Then  the $\sigma$ meson may be understood as a  
$(u\bar{u} + d \bar{d})/\sqrt{2}$ configuration having a definite 
mass $m_\sigma$ as predicted by the
quark-level NJL model.
The corresponding amplitude is constructed in analogy to the pseudoscalar pole
and given by
\begin{equation}
A^{t-{\rm pole}}_1(t)=\frac{g_{\sigma NN}F_{\sigma\gamma\gamma}}{
t-m^2_\sigma},
\label{sigmapole}
\end{equation}
with $g_{\sigma NN}$
being the $\sigma$-nucleon coupling constant, $F_{\sigma\gamma\gamma}$
the two-photon $\sigma$ decay amplitude and $m_\sigma$ the $\sigma$ 
mass\footnote{Occasionally it has been proposed to modify the pole amplitude
given in (\ref{sigmapole}) by considering the quantities 
$g_{\sigma NN}$ and $F_{\sigma\gamma\gamma}$ as $t$-dependent formfactors.
Such a procedure, however, is not allowed because it is incompatible with the
requirements of dispersion theory.}
. Then some consideration shows 
that the $t$-channel part of the  polarizability difference is given by
\begin{equation}
(\alpha-\beta)^t_{p,n}=\frac{g_{\pi NN}F_{\sigma\gamma\gamma}}
{2\pi m^2_\sigma}=\frac{5\alpha_eg_{\pi NN}}{6\pi^2 m^2_\sigma f_\pi}
=15.2\approx \frac{5\alpha_e g^2_{\pi NN}}{6\pi^2m^2_\sigma g_A {m_N}},
\label{alpha-beta-quarks}  
\end{equation}
in units of $10^{-4}{\rm fm}^3$, where $\alpha_e=1/137.04$, $g_{\sigma
  NN}\equiv g_{\pi NN}= 13.169\pm 0.057$ \cite{bugg04}, $f_\pi=(92.42\pm0.26)$
  MeV, 
 $m_\sigma= 666$ MeV as derived in subsection \ref{subsection2-2}
using the method of \cite{delbourgo95}. 
In (\ref{alpha-beta-quarks}) the
quark-model prediction
$F_{\sigma\gamma\gamma}=+|M(\sigma\to\gamma\gamma)|$
with $N_c=3$ has been used. The identity $g_{\sigma NN}\equiv g_{\pi NN}$
is predicted by the  L$\sigma$M and has been
experimentally confirmed by Durso et al. \cite{durso80}.
On the r.h.s. of Eq.
(\ref{alpha-beta-quarks}) use has been made of the approximately valid
Goldberger-Treiman relation where $g_A=1.255\pm 0.006$ is the axial vector
coupling constant. 

An upper limit for the  possible correction to $(\alpha-\beta)^t_{p,n}$
as given in (\ref{alpha-beta-quarks}) due to the $f_0(980)$ meson 
can be calculated
making the assumption that the coupling constants $g_{f(980) NN}$ and
$g_{\pi NN}$ are equal to each other. The result obtained is a possible
correction  of not more than $9 \%$.

\section{The quantity $(\alpha-\beta)^t$ predicted from the reactions
$\pi\pi\to\sigma\to\pi\pi$, $\gamma\gamma\to\sigma\to\pi\pi$ 
and $N\bar{N}\to\sigma\to\pi\pi$}

A different approach to a calculation of $(\alpha-\beta)^t$ which only makes 
use of experimental data without specific assumptions about the internal
structure of the $\sigma$ meson
is provided by the BEFT sum rule \cite{bernabeu74}. 
Instead of exploiting Eq.
(\ref{alpha-beta-quarks}) the
following relation is considered
\begin{equation}
\gamma\gamma\to|\sigma\rangle\to \pi\pi \to |\sigma\rangle
\to N\bar{N},
\label{eq-SUM3}
\end{equation}
and replaced by 
\begin{eqnarray}
&&\gamma\gamma\to|\sigma\rangle\to \pi\pi, \label{eq-SUM4a}    \\
&&\pi\pi \to |\sigma\rangle\to N\bar{N},
\label{eq-SUM4b}
\end{eqnarray}
obtained by means of a $t$-channel cut (see Fig. \ref{poleandcutgraphs} c)).
The further procedure is to use the 
unitarity relation
\begin{equation}
{\rm Im}_t T(\gamma\gamma\to N\bar{N})=\frac12 \sum_n
(2\pi)^4\delta^4(P_n-P_i)\, T(\gamma\gamma\to n)\, T^*(N\bar{N}\to n),
\label{eq-SUM5}
\end{equation}
where the sum on the right-hand side is taken over all allowed intermediate
states $n$ having the same total 4-momentum as the initial state. Furthermore,
$\pi\pi$ intermediate states are taken into account where the spin is
$J=0$ and the isospin $I=0$. These are the quantum numbers of the intermediate
$\sigma$ meson. 

If we restrict ourselves in the calculation
of the $t$-channel absorptive part to intermediate states with two
pions with angular momentum $J\leq 2$, the BEFT  sum rule \cite{bernabeu74}
gets a  convenient form for calculations:
\begin{eqnarray}
(\alpha-\beta)^t_{p,n}&=& 
 \frac{1}{16 \pi^2}\int^\infty_{4 m^2_\pi}\frac{dt}{t^2}\frac{16}{4{m_N}^2-t}
\left(\frac{t-4m^2_\pi}{t}\right)^{1/2}\Big[f^0_+(t)
  F^{0*}_{0}(t)\nonumber\\
&&-\left({m_N}^2-\frac{t}{4}\right)\left(\frac{t}{4}-m^2_\pi\right)
f^2_+(t) F^{2*}_{0}(t)\Big],\label{BackSR}
\end{eqnarray}
where
$f^{(0,2)}_+(t)$ and $F^{(0,2)}_0(t)$ are the partial-wave helicity
amplitudes of the processes $N\bar{N}\to \pi\pi$ and 
$\pi\pi\to \gamma\gamma$ with angular momentum $J=0$ and $2$,
respectively, and isospin $I=0$.

Except for the quantum numbers, properties of the $\sigma$ meson enter
through the amplitudes $f^{(0,2)}_+(t)$ and $F^{(0,2)}_0(t)$
corresponding to the
reactions (\ref{eq-SUM4a}) and (\ref{eq-SUM4b}). These amplitudes
incorporate the phase   $\delta^I_J(s)=\delta^0_0(s)$ which is extracted from 
$\pi$-$\pi$ scattering data as obtained in $\pi N\to N \pi\pi$ scattering
experiments. The information contained in the phase  $\delta^0_0(s)$ 
can be understood as being
due to a resonance with the pole parameters
\cite{eidelman04,colangelo01,ishida03}
\begin{equation}
\sqrt{s^{\rm pole}}\approx (500 -i\, 250)\,{\rm MeV}
\label{eq-SUM6}
\end{equation}
and the special property of  a $90^\circ$ crossing of the phase  at
\begin{equation}
\sqrt{s}(\delta^0_0=90^\circ)\approx 900 \,{\rm MeV}.
\label{eq-SUM7}
\end{equation}
The apparent mismatch of the $90^\circ$ phase crossing expected for the
resonance part as given by (\ref{eq-SUM6}) and the observed 
$90^\circ$ phase crossing as given by  (\ref{eq-SUM7})
has led to numerous considerations
\cite{ishida96,kaloshin04,bugg03}
among which the possible existence of a background 
\cite{ishida96} or a second pole at
negative mass parameter  $m^2$  
\cite{kaloshin04} or one pole with an $s$-dependent width
$\Gamma(s)$ \cite{bugg03} played a role.
This discussion shows that a generally accepted  model
for the $\sigma$ meson  as a resonant
state is still missing ( see, however, the last entry in  \cite{beveren86}
and references therein). Furthermore, the inclusion of the $\pi\pi$
phase relation into the $\gamma\gamma\to\pi\pi$ and $\bar{N}N\to \pi\pi$
amplitudes is not an easy task. This may be the reason that only recently 
some consistency in the prediction of $(\alpha-\beta)^t$ 
from the BEFT sum rule \cite{bernabeu74} has been obtained.
This has led to 
\begin{equation}
(\alpha-\beta)^t_{p,n}= (14.0\pm 2.0)\,\, 
\mbox{(Ref.\cite{schumacher05b})},
\,\,\,\,\, 16.46\,\, \mbox{(Ref. \cite{drechsel03})},
\label{BEFT}
\end{equation}
where the first value has been obtained by Levchuk et al.
(see\cite{schumacher05b}) and 
the second
value by Drechsel et al. \cite{drechsel03}.
The large error given in Ref. \cite{schumacher05b}
takes care of the uncertainties contained in
the appropriate choice of experimental data. 
The arithmetic average of the two results is 
$(\alpha-\beta)^t_{p,n}= 15.3\pm 1.3$.
This number has to be compared 
with the experimental data $(\alpha-\beta)^t_p=15.1\pm 1.3$,
$(\alpha-\beta)^t_n=14.8\pm 2.7$ (see Ref. \cite{schumacher05b}). 
We see that the predictions for $(\alpha-\beta)^t$ obtained from the
$\sigma$-meson pole based on the NJL model and from the BEFT sum rule
lead to agreement with each other and to agreement with
experiment.

\section{Discussion}

According to our
recent analysis \cite{schumacher05a,schumacher05b,levchuk05} the experimental 
polarizabilities may be summarized in the
form given in Table \ref{tablepolarizabilities}.
\begin{table}[h]
\caption{Summary on electromagnetic polarizabilities 
in units of $10^{-4}{\rm fm}^3$}
\begin{center}
\begin{tabular}{llll}
1&&\,\,\quad\quad\quad  proton& \,\,\quad\quad\quad  neutron\\
\hline
2&BL sum rule & $(\alpha+\beta)_p=13.9\pm 0.3$&$
(\alpha+\beta)_n=15.2\pm 0.5$\\
3&Compton scattering&$(\alpha-\beta)_p=10.1\pm 0.9$&
$(\alpha-\beta)_n=9.8\pm 2.5$\\
4&BEFT sum rule&$(\alpha-\beta)^s_p=-5.0\pm 1.0$&$(\alpha-\beta)^s_n=-5.0\pm
1.0$\\ 
5&line 3 -- line 4&
 $(\alpha-\beta)^t_p=15.1\pm 1.3$& $(\alpha-\beta)^t_n=14.8\pm
2.7$\\ 
\hline
6&experiment& $\alpha_p=12.0\pm 0.6$ 
&  $\alpha_n=12.5  \pm 1.7$ \\
7&$s$-channel only & $\alpha^s_p= \,\,\,4.5\pm 0.5$ 
&  $ 
\alpha^s_n=\,\,\,5.1\pm 0.6$\\
8&$t$-channel only &   $\alpha^t_p= \,\,\,7.5\pm 0.8$ 
&  $ 
\alpha^t_n=\,\,\,7.4\pm 1.8$\\
\hline
9&experiment& $\beta_p=\,\,\,1.9\mp 0.6$ 
& $\beta_n=\,\,2.7\mp 1.8$ \\
10&$s$-channel only& $\beta^s_p= \,\,\,9.5\pm 0.5$ & 
 $ 
\beta^s_n=10.1\pm 0.6$\\
11&$t$-channel only& $\beta^t_p= \,\,\,-7.6\pm 0.8$ & 
 $ 
\beta^t_n=-7.4\pm 1.9$\\
\hline
\end{tabular}
\end{center}
\label{tablepolarizabilities}
\end{table}
The quantities
$\alpha_p,\beta_p,\alpha_n,\beta_n$ are the experimental electric and magnetic
polarizabilities  for the proton and neutron, respectively. The 
quantities with an upper label $s$ are the corresponding electric and magnetic
polarizabilities where only the $s$-channel degrees of freedom are included. 
These latter quantities have been obtained by making use
of the fact that $(\alpha+\beta)$, when calculated from forward-angle
dispersion theory as given by the Baldin or Baldin-Lapidus (BL)
sum rule \cite{baldin60}
\begin{equation}
(\alpha+\beta)=\frac{1}{2\pi^2}\int^\infty_{m_\pi+\frac{m^2_\pi}{2{m_N}}}
\frac{\sigma_{\rm tot}(\omega)}{\omega^2}d\omega
\label{BLsumrule}
\end{equation}
has no $t$-channel contribution, i.e. $(\alpha+\beta)=(\alpha+\beta)^s$,  
and by using
the estimate $(\alpha-\beta)^s_{p,n}=-5.0\pm 1.0$ 
obtained form the $s$-channel part of the BEFT sum rule \cite{bernabeu74}
\begin{equation}
(\alpha-\beta)^s=\frac{1}{2\pi^2}\int^\infty_{m_\pi+\frac{m^2_\pi}{2{m_N}}}
\sqrt{1+\frac{2\omega}{m_N}}\left[\sigma(\omega,E1,M2,E3,\cdots)
-\sigma(\omega,M1,E2,M3,\cdots)\right]\frac{d\omega}{\omega^2}
\label{BEFTsumrule}
\end{equation}
both for the proton and the
neutron (see \cite{schumacher05b}).  The absence of a $t$-channel contribution
to $(\alpha+\beta)$ follows from the observation \cite{schumacher05b} that the
BL sum rule is fulfilled. This observation is explained by the fact
that in the forward direction vector-meson dominance converts the 
$t$-channel contribution into the Regge part of the photoabsorption cross
section. The numbers in  line 5
of Table \ref{tablepolarizabilities} are the $t$-channel contributions to 
$(\alpha-\beta)$ obtained from the experimental values 
$(\alpha-\beta)_p=10.1\pm 0.9$ and $(\alpha-\beta)_n=9.8\pm 2.5$ and the
estimate for    $(\alpha-\beta)^s_{p,n}$. As noted before, we see that the 
experimental values for
$\alpha$ are much larger than the $s$-channel contributions alone, whereas for
the magnetic polarizabilities the opposite is true. For the magnetic
polarizability it makes sense to identify the large difference between
the experimental value and the $s$-channel contribution with the diamagnetic
polarizability. This means that we may consider $\beta^t$ as the diamagnetic
polarizability.

Certainly, by identifying the expression obtained for
$(\alpha-\beta)^t_{p,n}$
in Eq. (\ref{alpha-beta-quarks}) with that of Eq. (\ref{BackSR}) 
a sum rule is obtained. This finding is also supported by the two graphs
$b)$ and $c)$ in Fig. \ref{poleandcutgraphs}. Furthermore, we see
in Table \ref{summary} that the experimental results obtained
for $(\alpha-\beta)^t$ from the  experiments, from the $\sigma$-meson
pole and from the $\sigma$-meson cut agree with each other and thus give 
also strong
support for the existence and validity of the sum rule.  
\begin{table}[h]
\caption{Difference of  
electromagnetic polarizabilities $(\alpha-\beta)^t_{p,n}$ in the $t$-channel
(in units of $10^{-4}{\rm fm}^3$). The result given for the $\sigma$-meson
cut or BEFT sum rule is the arithmetic average of the results of
Drechsel et al. \cite{drechsel03} and Levchuk et al. 
(see \cite{levchuk05,schumacher05b})}
\begin{center}
\begin{tabular}{@{}lll@{}}
\hline
&$(\alpha-\beta)^t_p$ &   $ (\alpha-\beta)^t_n$         \\
\hline
experiment&$15.1 \pm 1.3$& $14.8 \pm 2.7$\\
$\sigma$-pole& $15.2$&   $15.2$                    \\
BEFT sum rule& $15.3\pm 1.3$&  $15.3\pm 1.3$         \\
\hline
\end{tabular}
\end{center}
\label{summary}
\end{table}
We expect that by studying this sum rule in more
detail some more insight into the structure
of the $\sigma$ meson is obtained. In such a study the role of the 
quark-level NJL model would be to describe the $\sigma$ meson as the
particle of the $\sigma$ field
with a definite mass $m_\sigma =666$ MeV, whereas the BEFT sum rule
exploits  on-shell properties of the $\sigma$ meson as there is  {\it e.g.} 
the phase relation $\delta^0_0(s)$.  The sum rule we are proposing
has the property of linking on-shell aspects of the $\sigma$ meson with
properties of the $\sigma$ meson as the particle of the $\sigma$ field.

As a concluding remark we wish to state that the present paper closes
a circle, starting with the work of Hearn and Leader (1962) \cite{hearn62} 
where the role of the $t$-channel contributions to the Compton scattering
amplitudes has been clarified. In the present work we have shown that the
$(\sigma,\pi^0)$  particle doublet is capable of quantitatively reproducing
this $t$-channel contribution, with  minor additional contributions from 
the $f_0(980)$, the $\eta$ and the $\eta'$ meson. Since the
$t$-channel contribution cannot be interpreted in terms of nucleon
resonances or in terms of the meson cloud as viewed in photon-meson
production  experiments, it is reasonable to interpret this contribution
in terms of a short-range property of the constituent quarks, as proposed in
\cite{schumacher05b}. More explicitly this means that each constituent
quark $|q\rangle$ of the nucleon is converted into a 
$|(\sigma,\pi^0)q\rangle$  $t$-channel resonant intermediate
state during the Compton scattering process. However, 
this resonant intermediate state is located in the unphysical 
region at positive $t$. 
At $\theta=\pi$ the $\pi^0$ meson is involved in those scattering processes
where the incoming and the outgoing photon have perpendicular directions of
linear polarization whereas for the $\sigma$ meson the directions of linear
polarization are parallel.

Other approaches using model calculations 
(see \cite{schumacher05b} for a summary)  or diagrammatic techniques
\cite{bernard91,hildebrandt04,hemmert98} do not take
care of the scalar-isoscalar $t$-channel contribution in a sufficient way. 
This means that they are important with respect to a test of the underlying
theoretical ansatz, but they
cannot be considered as a
quantitative descriptions of the electromagnetic 
polarizabilities of the
nucleon. It is very interesting to analyze  the different strategies 
contained in these approaches \cite{bernard91,hildebrandt04,hemmert98}
to cope 
with the serious problem caused by the missing scalar-isoscalar 
$t$-channel in the light of dispersion theory. 
The most transparent of these strategies  are the neglect
of the contribution of the $\Delta$ resonance \cite{bernard91}
and the introduction
of an empirical counter term of unnatural size   \cite{hildebrandt04}.

The strategy of neglecting the $\Delta$ resonance contribution
has been analysed by L'vov \cite{lvov93} in terms of dispersion theory.
It has been confirmed that the simultaneous neglect of the scalar-isoscalar
$t$-channel and of the $\Delta$ resonance leads to an approximate agreement
with the experimental data. 
Furthermore, it has been confirmed that the 
use of the heavy baryon approximation ($m_N\to \infty$) leads to an improvement
of the agreement with the experimental data, though no 
good physical reason has been found for such a replacement. Finally, it 
has been shown \cite{lvov93} that the evaluation of chiral
loops  \cite{bernard91} leads to an approximate agreement with  the
contributions of the 
$E_{0^+}$ component to  $\alpha$ and $\beta$. 
However, this is only the case when the
empirical CGLN amplitude $E_{0+}$ is replaced by the Born approximation 
 $E^{\rm Born}_{0+}$.

Since there is no
definite interpretation available for the empirical counter terms of 
unnatural sizes \cite{hildebrandt04}, $\delta\alpha$ and $\delta\beta$,
it may be allowed to tentatively compare these terms with the   
present $t$-channel polarizabilities $\alpha^t_{p,n}$ 
and $\beta^t_{p,n}$ contained in
Table \ref{tablepolarizabilities}.
Qualitatively, these terms $\delta\alpha$ and $\delta\beta$ 
are introduced to produce diamagnetism. Furthermore, their physical nature
is assumed to be related to short-distance phenomena of some kind. These two
properties suggest that    $\delta\alpha$ and $\delta\beta$ on the one hand 
and $\alpha^t_{p,n}$ and $\beta^t_{p,n}$ on the other should have some common
features or even may be identical. Unfortunately, the numbers obtained
empirically,  i.e. 
$\delta\alpha=-5.92\pm 1.36$ and $\delta\beta=-10.68\pm 1.17$, are not in a
good agreement with the numbers obtained for  
$\alpha^t_{p,n}=-\beta^t_{p,n}=7.6$.

\section{Summary}

The present paper succeeds in a quantitative derivation of
the $\sigma$-meson pole contribution to $(\alpha-\beta)^t$
which formerly was introduced and 
treated semi-quantitatively by adjusting to experimental data \cite{lvov97}. 
Furthermore,
there was no strict argument for the pole-structure of this  contribution,
since the only knowledge about the scalar-isoscalar $t$-channel
was obtained from the BEFT sum rule which makes use of a $t$-channel
cut and not of a $t$-channel pole. In the present paper we show that
all the relevant parameters of the $\sigma$-meson pole, {\it viz.} the
$\sigma$-meson mass $m_\sigma$ and properties of the two-photon decay width
$\Gamma_{\gamma\gamma}$ follow from the NJL model without any adjustable
parameter. Making use of previous work \cite{hatsuda94,delbourgo95}, 
the mass $m_\sigma$ is  obtained in two different versions
of the NJL model and two different
regularization schemes. In the first case \cite{hatsuda94} the four-fermion
version of the NJL model is used and the regularization scheme makes use of
a cut-off parameter $\Lambda$. Both quantities, the cut-off parameter
$\Lambda$ and the coupling $G$ are obtained within this regularization
scheme by adjusting to empirical data as there are the pion decay constant
$f_\pi$, the pion mass $m_\pi$ and the current-quark mass $m_0$. The result
is $m_\sigma=668$ MeV.
In the second case \cite{delbourgo95} the bosonized version of the NJL model
is applied and the 
regularization scheme makes use of dimensional scaling, {\it i.e.} no
cut-off parameter $\Lambda$ is present so that the integrals extends to
infinity. Instead, use is made of the fact that the difference of the diverging
integrals for the constituent-quark mass $M$ and the pion decay constant 
$f_\pi$ is finite. The result is $m_\sigma=666$ MeV. It is remarkable and
has not been pointed out before that these two completely 
different treatments
of the NJL model  perfectly lead to the same result. This gives us
confidence  that conclusions based on this number for the $\sigma$
meson mass are on a sound basis. 
In case of the two-photon decay width $\Gamma_{\gamma\gamma}$ the
NJL model is used to justify the neglect of couplings of the $\sigma$ meson
to two photons via meson loops. Such additional couplings follow from the
L$\sigma$M. If taken into account the additional couplings would partly destroy
the good agreement between theory and experiment. Therefore, the argument
delivered by the NJL model is important. 
     
A further new results obtained 
is  the calculation of the two-photon decay width
$\Gamma_{\gamma\gamma}$ of the $f_0(980)$ meson. This calculation is based
on a $s\bar{s}$ structure of the $f_0(980)$ meson and leads to
agreement with  experimental results. We consider this as an argument that
indeed the  $f_0(980)$ meson and the $\sigma$ meson have a $q\bar{q}$ core
and not a $qq\bar{q}\bar{q}$ structure as suggested in other approaches
\cite{jaffe76}. 
This result is important for the present investigation 
as well as for the physics
of scalar mesons in general. In addition, for the first time the contribution
of the $f_0(980)$ meson to $(\alpha-\beta)^t$ has been calculated. 
A further new
result are arguments in favor of positive decay amplitudes
$F_{\eta\gamma\gamma}$ and $F_{\eta'\gamma\gamma}$ which formerly were
believed to be negative.

Via the two representations of $(\alpha-\beta)^t$ a
quantitative link is obtained between the $\sigma$ meson as the particle
of the $\sigma$ field and the $\sigma$ meson as showing up as an extremely
shortlived intermediate state in reactions where two pions are involved.
This observation may be exploited to get insight into the structure of the 
$\sigma$ meson by constructing a model which makes the relation between the
graphs b) and c) in Figure \ref{poleandcutgraphs} transparent.

\section*{Acknowledgment}
The author is indebted to Deutsche Forschungsgemeinschaft for the support of
this work through the projects SCHU222 and 436RUS113/510. He thanks
M.I. Levchuk, A.I. L'vov and A.I. Milstein for their continuous
interest in this work and for many comments and suggestions.  He also 
thanks E. van Beveren, F. Kleefeld, G. Rupp, and M.D. Scadron for many useful
comments regarding their work.

\end{document}